\documentclass{article}
\usepackage[utf8]{inputenc}
\usepackage{caption}
\usepackage{lmodern}
\usepackage{amsmath}
\usepackage{aas_macro}
\usepackage[margin=.8in]{geometry}
\usepackage[pagebackref,colorlinks = true, urlcolor  = blue, linkcolor = blue, citecolor = blue, anchorcolor = black]{hyperref}
\renewcommand*\backref[1]{\ifx#1\relax \else [cited in page(s) #1] \fi}
\usepackage{orcidlink}
\DeclareMathOperator{\dif}{d\!}

\DeclareMathOperator{\e}{e\!}
\def\mchi{m_{\rm DM}}
\def\sv{\langle\sigma v\rangle}
\def\mq{\mathcal Q}
\def\mr{\mathcal R}
\def\ms{\mathcal S}
\def\mf{\mathcal F}
\def\mm{\mathcal M}
\def\mn{\mathcal N}
\def\diffsph{{\tt diffSph}}

\def\jeq{{\color{jupyterpurple}\fontseries{b}\selectfont =} }
\definecolor{jupytergreen}{rgb}{0, .502, .051}
\definecolor{jupyterpurple}{rgb}{.37, .13, .50}
\definecolor{jupyterblue}{RGB}{64,128,128}
\definecolor{jupyterred}{RGB}{186,34,32}

\title{diffSph: a {\tt Python} tool to compute diffuse signals from dwarf \\ spheroidal galaxies}
\author{Martin Vollmann\orcidlink{0000-0003-2193-8922}, Finn Welzm\"uller\orcidlink{0009-0009-2456-8884}, Lovorka Gajovi\'c\orcidlink{0000-0001-9512-8565}}
\date{}

\begin{document}
\maketitle

\begin{abstract}
    So far no diffuse emissions in dwarf spheroidal satellites of the Milky Way have ever been observed.
    Given that dwarf galaxies are predominantly composed of Dark Matter, the discovery of these signals could offer valuable insights into understanding the nature of Dark Matter.
    We present ``diffSph'', a Python tool which in its present version provides fast predictions of such diffuse signals in radio frequencies.
    It also features a very comprehensive module for the computation of ``J'' and ``D'' factors that are relevant for indirect Dark Matter detection using gamma rays.  
    Routines are coupled to parton-shower algorithms and Dark Matter halo mass functions from state-of-the-art kinematic fits.
    This code is also useful for testing generic hypotheses (not necessarily associated with any Dark Matter candidate) about the cosmic-ray electron/positron sources in the dwarf galaxies.
    The diffSph tool has already been employed in searches for diffuse signals from dwarf spheroidal galaxies using the LOw Frequency ARray (LOFAR). 
\end{abstract}

\section{Introduction}
Due to their small star formation rates, Milky Way satellite dwarf spheroidal galaxies (dSph) are weak emitters, and to this date, no diffuse emission from these objects has been observed at any frequency. 
Most of their scarce luminosity is produced by their member stars instead.
Dwarf galaxies are, however, Dark-Matter (DM) rich environments and in several theories of particle DM such diffuse emissions are predicted via annihilation or decay \cite{Zeldovich:1980st}.
While the signal strengths are typically weak, they are typically within the detectable range of current telescopes.
Several collaborations have already conducted dedicated searches for these signals using e.~g. gamma-ray telescopes \cite{Ahnen:2016qkx} and found no compelling evidence of any diffuse signal in the observed targets.
Instead, their non-detection results have been used in order to set exclusion limits on the parameter space of particle DM models.

Gamma rays are the ``golden'' messenger for indirect DM detection better-controlled uncertainties, theoretically motivated models (e.~g. WIMP), phenomenologically promising signatures and ever increasing improvements in sensitivity.
Nevertheless, complementary indirect searches using radio interferometers such as the LOw Frequency ARray (LOFAR) can yield to even more stringent exclusion bounds on particle DM models \cite{2020arXiv201111947V,Gajovic:2023bsu}.
The latter limits are, however, subject to much larger uncertainties than those encountered in the gamma-ray counterpart. 

Independent of these intrinsic limitations, if diffuse synchrotron-radiation emission in dSphs is discovered soon, solid theoretical interpretations are necessary.
Therefore, we introduce \diffsph{}, a {\tt Python} tool that produces signal templates for diffuse synchrotron emissions within the context of diffusive cosmic-ray propagation powered by magnetic-field turbulence and for \emph{generic} hypotheses on the origin of these signals.
Special focus is put on the hypothesis that the cosmic-ray electrons are produced via DM annihilation or decay but the modularity of \diffsph{} allows for several other hypotheses (not associated to DM) to be tested.

More specifically, the \diffsph{} package includes several pre-computed objects that speed up the computations that are relevant for testing several interesting/popular particle DM scenarios.
For instance, \diffsph{} can efficiently generate synchrotron spectra from decaying or WIMP DM with particle masses in the 1~GeV - 100~TeV range by evaluating pre-computed kernel functions. 
This renders its performance quite efficient without losing much accuracy.

The \diffsph{} code has also a module (\texttt{diffsph.profiles}) that computes the line-of-sight and angular integrals that are relevant for indirect DM detection phenomenology in general.
This is linked to several mass models of dwarf galaxies found in the literature.
In particular, this module uses these mass models as input in its routines for the computation of the so-called astrophysical $J$- and $D$-factors, which are essential pieces for indirect DM detection with gamma rays.
The code can be easily accessed and downloaded via  \href{https://github.com/mertio1/diffsph}{GitHub}, e.~g.
\bigskip

\noindent{\tt git clone https://github.com/mertio1/diffsph.git}
\bigskip

If the user wants to install \diffsph{} globally, they can do it by going into \diffsph{}'s main folder and on a terminal typing (the installation can take a few minutes)
\bigskip

\noindent\texttt{python setup.py bdist\_wheel}

\noindent{\tt pip install .}
\bigskip

An extensive online documentation is available at \href{https://mertio1.github.io/diffsph/}{https://mertio1.github.io/diffsph/} and, as a PDF file, at this \href{https://raw.githubusercontent.com/mertio1/diffsph/master/documentation.pdf}{URL}. 
The main novelty of \diffsph{} with respect to already existing tools (e.~g.  \cite{McDaniel:2017ppt,Hutten:2018aix}), is that instead of using a method-of-images expansion for the cosmic-ray electron trasport equation, a Fourier decomposition is employed which turns out to be more suitable for the computation at hand. 

The article is organized as follows: in section \ref{sec:theory} we provide a review about the physical modelling of the diffuse emission in dwarf galaxies.
In the next part (section \ref{sec:arch}) we describe how \diffsph{} is structured and explain how the different hypotheses that are incorporated in the code can be tested using it.
In section \ref{sec:results} we show some concrete examples and we then conclude.
In the appendices we include a brief discussion on some asymptotic approximations and analytical results that are adopted by \diffsph{}.

\section{Theoretical background}
\label{sec:theory}
The description of the transport of cosmic-ray electrons (CREs) in dSphs is a puzzling problem.
This is because \emph{no} magnetic fields in these galaxies have been observed and the resolution of existing cosmological magneto-hydrodynamic simulations is not sufficient in order to infer their distribution and strength in these galaxies.
In this work we adopt the arguably simplest transport model for the CRE, which has become the standard lore in the literature, see Ref. \cite{Colafrancesco:2006he} and references therein. 
In this framework, the CRE number density $n_e(r,E)$ as a function of the galacto-centric radius $r$ and the CREs' energy $E$, is obtained by solving a stationary and spherically-symmetric diffusion-loss equation 
\begin{equation}
\label{eq:diffloss}
	D(E)\nabla^2n_e+\frac{\partial}{\partial E}[b(E)n_e]+q_e=0\ ,
\end{equation}
and a boundary condition $\left.n_e(r,E)\right|_{r=r_h}=0$, where $r_h$ is the \emph{visible halo radius} parameter. 
The model is an adaptation of the Milky Way's leaky-box CR transport model to the geometry of the dwarf spheroidal galaxies.
In particular, the diffusion and energy-loss coefficients are respectively given by $D(E)=D_0(E/E_0)^\delta$ and $b(E)= b_0\,[1+(B/B_c)^2] E^2$, where $\delta$, $D_0/E_0^\delta$, $b_0$, $B_c$ are constants\footnote{We use 
$e=4.80326\times10^{-10}$~Fr, $m_e=9.1093837\times10^{-28}$~g and $T_\textrm{CMB}=2.725$~K. } and $B$ is an \emph{effective parameter} that captures the effect of the (randomly oriented) magnetic field in the galaxy.
It is assumed that $B$ is constant within the $r<r_h$ volume.

The energy-loss coefficient receives contributions from two processes: synchrotron and inverse Compton scattering (ICS) radiation.
The former is proportional to the squared magnetic field $B$ and for the latter we neglect all radiation fields but the cosmic microwave background (CMB):
\begin{eqnarray}
    b_0 &=& \frac{32\pi^3e^4k_B^4T_\textrm{CMB}^4}{135\hbar^3m_e^4c^{10}}\simeq2.652\times10^{-17}\,\textrm{GeV}^{-1}\textrm{s}^{-1}\ , \quad (\texttt{utils.consts.b0})\ ,\\
    B_c &=& \sqrt{\frac{8\pi^3k_B^4T_\textrm{CMB}^4}{15\hbar^3c^3}}\simeq3.238\,\mu\textrm{G}\ ,\quad (\texttt{utils.consts.Bc})\ ,
\end{eqnarray}
the bracketed variable names in the equations above indicate their corresponding nomenclature in \diffsph{}.

The source term $q_e(r,E)$ in Eq. \eqref{eq:diffloss} can be factored as 
\begin{equation}
q_e(r,E)=\mq\, \mr(r)\,\ms(E)\ ,
\end{equation}
where $\mr(r)$, $\ms(E)$ are hypothesis-dependent functions; and $\mq$ is a normalization constant. 
We consider generic hypotheses, e.~g. WIMP, decaying DM, point source, etc. (see Table \ref{tab:modules}). 
In the WIMP hypothesis case, for example, the normalization constant is given by\footnote{If the variable \texttt{self\_conjugate} is set \texttt{True}} $\mq=\sv/(2\mchi^2)$, where $\sv$ is the total velocity-averaged annihilation cross section times the relative WIMPs' velocity and $\mchi$ is their mass; $\mr(r)=\rho^2(r)$, where $\rho(r)$ is the DM halo density and $\ms(E)\equiv (1/\sv) \times \dif\, \sv / \dif E$ is the WIMP-model specific differential electron/positron ``yield'' per annihilation.

\begin{table}[!ht]
\centering 
\begin{tabular}{c||cccc}
Hypothesis & $\mq$ & $\mr(r)$ & $\ms(E)$ \\
\hline
\hline\\[-5pt]
\tt 'wimp' & $\sv/(2\mchi^2)$ & $\rho^2(r)$ & $(1/\sv) \times \dif\, \sv / \dif E$ \\[5pt]
\tt 'decay' & $\Gamma_\chi/\mchi$ & $\rho(r)$ & $(1/\Gamma_\chi)\times\dif\, \Gamma_\chi /\dif E$ \\[5pt]
\tt 'generic' & $\mn$ & $\mr_\star(r)$ & $E_0^{\Gamma-1}/E^\Gamma$ 
\end{tabular}
\caption{\label{tab:modules} Possible hypotheses of \diffsph{} and their associated source-term functions and parameters. 
} 
\end{table}

\subsection{Fourier-expanded solution}
Independent of the hypothesis, Eq. \eqref{eq:diffloss} can be solved semi-analytically as a linear combination of (Fourier) basis functions so that e.~g. the brightness $I_\nu(\theta)$ as a function of the observed angular radius $\theta$ can be expressed as 
\begin{equation}
\label{eq:bright}
    I_\nu(\theta) = \frac{\mq}{4\pi}\sum_{n=1}^\infty h_n\times X_n(\nu)\times f_n(\theta/\theta_h)
\end{equation}
where $R$ is the dSph's distance and the variable $\theta_h=\arcsin(r_h/R)\approx r_h/R$ is the angular radius of the diffuse signal. The basis functions $f_n(x)$, which are defined as
\begin{equation}
\label{eq:basisfuncs}
    f_n(x)=2\,\int_x^1\frac{\sin(n\pi y)\,{\rm d}y}{\sqrt{y^2-x^2}} \ ,
\end{equation}
do not depend on either the CRE sources nor their propagation: they are {\it universal} \cite{2020arXiv201111947V}.

The remaining $\mq\times h_n\times X_n(\nu)$ terms in the formula carry all the dependence on both the transport and CRE source models. 
Concretely, the $n$-th order \emph{halo/bulge factors} $h_n$ are given by
\begin{equation}
\label{eq:hfac}
    h_n=\frac2{r_h}\int_0^{r_h}{\rm d}r r\sin\frac{n\pi r}{r_h}\mr(r) \ .
\end{equation}
and the \emph{spectral coefficients} $X_n(\nu)$ can be written in terms of the spectral function $X(\nu)$ as $X_n(\nu)\equiv X(\nu;\tau_0/n^2)$.
The latter is given in terms of the {\it diffusion time scale} parameter $\tau_0=r_h^2/D_0$ by
\begin{equation}
\label{eq:spec}
X(\nu;\tau_0)=\frac{2\sqrt3e^3B}{m_ec^2}\int_0^\infty\frac{{\rm d}z}z\mf(z)\frac{E_c(\nu/z)}{b(E_c(\nu/z))}{\rm  e}^{-\eta(E_c(\nu/z);\tau_0)}\int_{E_c(\nu/z)}^\infty{\rm d}E\,\ms(E){\rm e}^{+\eta(E;\tau_0)}\ ,
\end{equation}
where $E_c(\nu)=\sqrt{2\pi m_e^3c^5\nu/(3eB)}=\epsilon_0\sqrt{\nu/B}$ ($\epsilon_0=5.576$~GeV~$\mu$G${}^{1/2}$~GHz${}^{-1/2}$, \texttt{utils.consts.eps0}) can be understood as the typical energy of a CRE emitting synchrotron radiation at frequency $\nu$ for a magnetic field $B$.

The \diffsph{} package includes several built-in templates for the radial functions $\rho(r)$ and $\mr_\star(r)$ and for the electron/positron yields $\ms(E)$ (see Tab. \ref{tab:modules}). 
While the radial templates $\mr(r)$ are given analytically in Tables \ref{tab:rhos}-\ref{tab:R(r)}, the $\ms(E)$ functions have to be computed numerically in the case of DM annihilation or decay.
In the standard approach, $\ms(E)$ is given by
\begin{equation}
    \ms(E)=\sum_{I}\textrm{BR}_I\frac{\dif N_I}{\dif E}\ ,
\end{equation}
where BR stands for \emph{branching ratio}, i.~e. BR${}_I=\sv_I/\sv$, ($\sum_I$BR${}_I=1$), and the index $I$ runs over primary final-state particle/antiparticle pairs (referred to as \emph{channels} in \diffsph{}). 
These depend on the mass of the DM particle and are readily available in specialized software tools like \texttt{DarkSUSY} \cite{Bringmann:2018lay}, \texttt{MicrOMEGAS} \cite{Belanger:2010gh} and PPPC \cite{Cirelli:2010xx}.
In this work, we use the (electroweak corrected) results by PPPC for the following possible primary final states: 
\begin{align}
\texttt{channel}=\{e^+e^-\ (\texttt{`ee'} \textrm{ or } \texttt{`ee\_MC'}), \mu^+\mu^-\ (\texttt{`mumu'}),{}&{} \tau^+\tau^-\ (\texttt{`tautau'}), \nu\bar\nu\ (\texttt{`nunu'}), \nonumber\\
q\bar q\ (\texttt{`qq'}), c\bar c\ (\texttt{`cc'}), b\bar b\ (\texttt{`bb'}), t\bar t\ (\texttt{`tt'}),  {}&{} W^+W^-\ (\texttt{`WW'}), ZZ\ (\texttt{`ZZ'}), hh\ (\texttt{`hh'})\}\ ,
\end{align}
where $q$ refers to any neutral combination of the light quarks $u$, $d$, $s$ and their antiparticles.
In the $e^+e^-$ case, there are two possibilities.
The \texttt{channel = `ee'} scenario corresponds to the $\ms(E)=\delta(E-\mchi)$ hypothesis, whereas \texttt{channel = `ee\_MC'} incorporates soft/collinear radiation effects which are obtained numerically using the aforementioned software packages.

\section{Architecture of \rm\diffsph{} }
\label{sec:arch}
\begin{figure}[th!]
    \begin{center}
    \includegraphics[width=.8\textwidth]{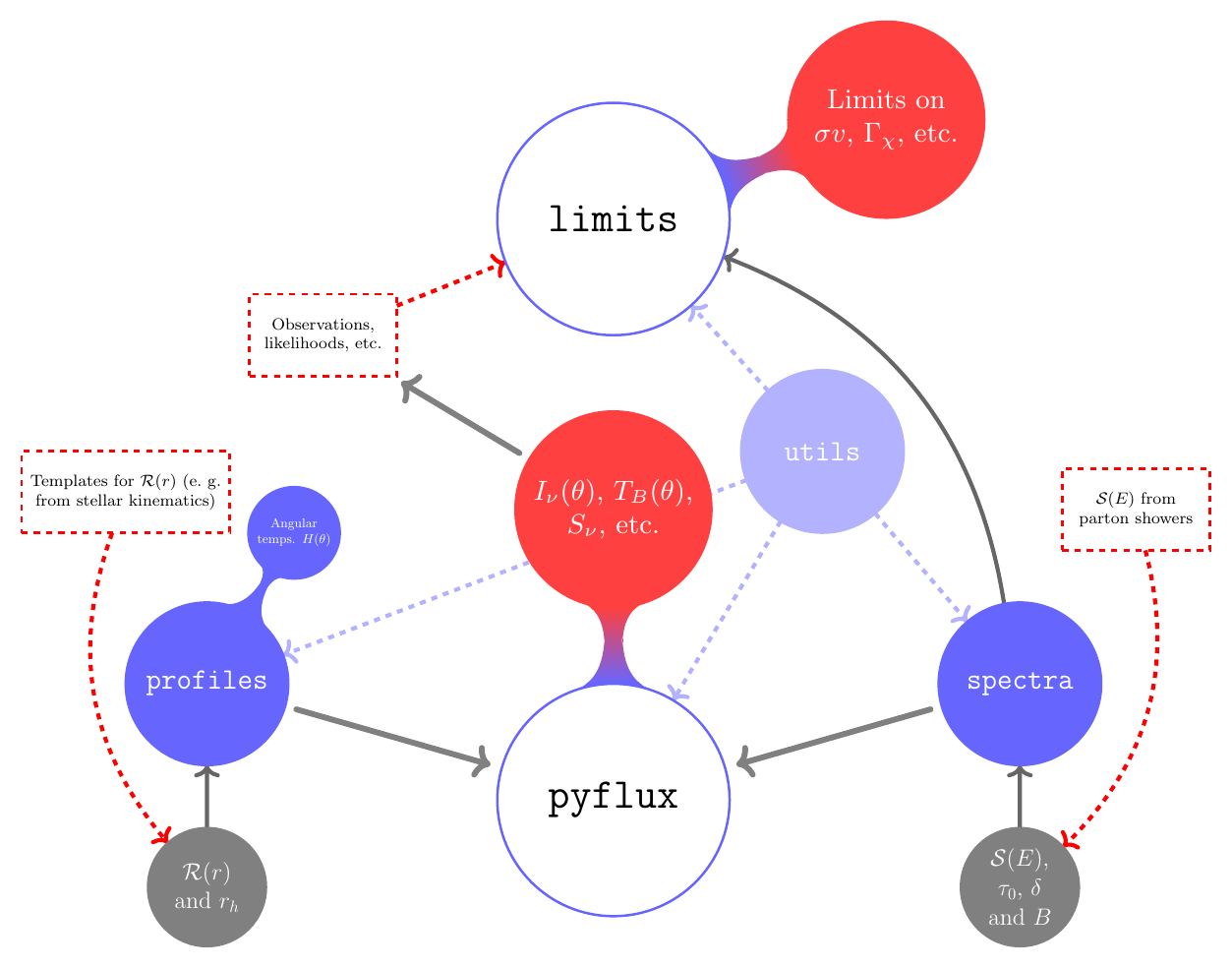}
    \caption{\label{fig:flowchart} Code architecture of \diffsph{}. Main modules are represented as white blobs with blue borders and their extensions (red blobs) show the associated top-level functions. Sub-modules are shown in blue while the input parameters in gray.
    Rectangular nodes correspond to outsourced inputs and their use is optional.}
    \end{center}
\end{figure}

Fig. \ref{fig:flowchart} shows what the main capabilities of \diffsph{\,1.0.} are and how these are interconnected.
The chief purpose of the code is obtaining fast predictions for synchrotron fluxes in dwarf galaxies once the hypothesis is given. 
This can be done using the \texttt{pyflux} module.
Users can also exclude hypotheses given some experimental data using the \texttt{limits} module.

\subsection{Predicted fluxes with \texttt{pyflux}}    
The \texttt{pyflux} module calls python functions that are contained in the \texttt{profiles} and \texttt{spectra} submodules in order to obtain diffuse emission fluxes for a given object and a set of parameters.
For example, if the user wants to obtain the brightness $I_\nu(\theta)$ of a given galaxy \texttt{galaxy}, at the frequency \texttt{nu} (in GHz) and as a function of angle \texttt{theta} (in arcmin) in the WIMP hypothesis, the \texttt{pyflux} module will compute and add all necessary series factors in Eq. \eqref{eq:bright} for that particular galaxy. 
The user has to specify which radial template $\texttt{rad\_temp}$ (Einasto, NFW, etc.) and the reference from which the relevant parameters ($\rho_s$, $r_s$, etc.) should be extracted. 
The current version includes mass models from Refs. \cite{2015MNRAS.451.2524M,Geringer-Sameth:2014yza,Ichikawa:2016nbi,Ichikawa:2017rph,Sanchez-Conde:2011zys} (see Table \ref{tab:reffits}).
Half-light radii and distances are obtained from \cite{McConnachie:2012vd,Martin:2008wj,2011AJ....142..128W}.

The default transport-model parameter set of \diffsph{} is inspired by CR propagation fits from the Milky Way such as \cite{Porter:2021tlr}: $B=2\,\mu$G, $D_0=2.77\times 10^{28}$~cm${}^2$/s and \footnote{However, note that other authors \cite{Korsmeier:2021bkw,Kahlhoefer:2021sha,Genolini:2021doh,Evoli:2019iih} obtain $\delta\simeq 0.5$ consistent with Kraichnan diffusion.} $\delta=0.35$. 
Note that these parameters are chosen for the sake of concreteness as no measurement of any of these exist and that the predictions that are obtained by using them are fairly optimistic. 
As a concrete example, the \diffsph{} prediction for the brightness (in Jy/sr) of a 50~GeV WIMP annihilating into $\mu^+\mu^-$ (i.~e. best-fit model for the Galactic-Center-Excess proposed in Ref. \cite{DiMauro:2021qcf}) at 150~MHz that is emitted from \textit{Segue 1} at $\theta=0$ and adopting a Herquist/Diemand/Zhao (HDZ) mass model with parameters fitted by Ref. \cite{Geringer-Sameth_2015}, is obtained by typing the following command:
\bigskip

\noindent
\begin{center}
\begin{tabular}{|ll|}
\hline \\ [-5pt]
\ttfamily \fontseries{l}\selectfont In[ ]:      &  \ttfamily {\color{jupytergreen}{\fontseries{b}\selectfont from}} {\fontseries{l}\selectfont diffsph} {\color{jupytergreen}{\fontseries{b}\selectfont import}} {\fontseries{l}\selectfont pyflux} {\color{jupytergreen}{\fontseries{b}\selectfont as}} {\fontseries{l}\selectfont pf} 
\\[5pt]
{} & \ttfamily \fontseries{l}\selectfont
pf.synch\_brightness(\\
{} & \ttfamily \fontseries{l}\selectfont\qquad theta \jeq {\color{jupytergreen} 0.}, nu \jeq {\color{jupytergreen} .15}, galaxy \jeq {\color{jupyterred} 'Segue I'}, rad\_temp \jeq {\color{jupyterred} 'HDZ'}, ref \jeq {\color{jupyterred} '1408.0002'},  \\
{} & \ttfamily \fontseries{l}\selectfont\qquad 
hyp \jeq {\color{jupyterred} 'wimp'}, sv \jeq {\color{jupytergreen} 3.e-26}, mchi \jeq {\color{jupytergreen} 50.}, channel \jeq {\color{jupyterred}  'mumu'}) \\ [10pt]
\ttfamily \fontseries{l}\selectfont Out[ ]:      & \ttfamily \fontseries{l}\selectfont 1.1055442452868214
\\[5pt]
\hline
\end{tabular}
\end{center}
\bigskip

\subsubsection{\boldmath Determination of \texorpdfstring{$N$}{N}}
The solution given in Eq. \eqref{eq:bright} is an infinite series on Fourier-like modes $n=0,1,2,\ldots$ 
However, \diffsph{} computes this up to a maximum value $n=N$.
In order to find a prescription for the determination of $N$ that is reasonably large, there are two aspects that need to be considered.

First, as usual in Fourier any decomposition, the larger the mode index $n$ the better the spatial resolution of the solution.
As a rule of thumb, $N\sim 1/\delta_\textrm{res}$ where $\delta_\textrm{res}\sim \Delta r/r \simeq \Delta\theta/\theta$ is the targeted angular (radial) resolution of the theoretical prediction \eqref{eq:bright}.
On the hand, regardless of the $r$- or $\theta$-resolution arguments, a much stronger condition that the solutions have to fulfill is that these should convergence on a global level.
More specifically, $N$ has to be large enough that the total flux density as integrated over the complete galaxy leads to a convergent result. 
This is given by
\begin{equation}
\label{eq:fd}
S_\nu = \frac1{R^2}\frac{r_h^2\,\mq}\pi\sum_{n=1}^\infty \frac{(-1)^{n-1}}n h_n\times X_n(\nu)+\mathcal O\left(r_h^4/R^4\right)\ .
\end{equation}

We thus introduce two variables, a Boolean one \texttt{high\_res}, which is set \texttt{False} by default and a floating one \texttt{accuracy} (in \%).
Given these two, \diffsph{} determines $N$ by evaluating the ratio $|s_{m+1}/S_m|$ for ever increasing values of $m$, where $s_{m}$ is defined as
\begin{equation}
s_m=h_m X_m \ , 
\end{equation}
and, depending on the \texttt{high\_res} variable, $S_m$ is given by
\begin{eqnarray}
    S_m=\left\{
    \begin{array}{c}
    \displaystyle
    \sum_{n=1}^m\frac{(-1)^{n-1}}ns_n\quad ,\quad \texttt{high\_res = True}\\
    \displaystyle
    \sum_{n=1}^m (-1)^{n-1}s_n\quad ,\quad \texttt{high\_res = False}
    \end{array}
    \right. \ .
\end{eqnarray}

The algorithm (\texttt{pyflux.which\_N}) then stops when the ratio is smaller than the chosen value (default value is 1~\%) of the \texttt{accuracy} variable. 
The value of $m$ at which the algorithm stops will be the chosen $N$.
A key feature of \diffsph{} is that while this algorithm is run, the code stores the values of all series coefficients $\{s_m\}_{m=1}^{N}$ in the cache folder, so that the computations are not carried out all over again when the same set of model parameters are inputted.

\subsection{The \texttt{limits} module}
Besides generating flux predictions via the \texttt{pyflux} module, \diffsph{} offers the possibility of producing 2$\sigma$ exclusion limits (or limit forecasts) on the relevant signal-strength parameters of any hypothesis (e.~g. $\sv$ in the case of the WIMPs) given a non-detection.
This is done by the \texttt{limits} module. 
There are two possibilities which we illustrate (in the context of the WIMP hypothesis) below.

\subsubsection{\texttt{sigmav\_gausslim}}
This is the strategy that was employed in Refs. \cite{Gajovic:2023bsu,Vollmann:2019boa}.
In this approach, we use the fact that the leading-mode shape function $f_{n=1}(\theta/\theta_h)$ defined in \eqref{eq:basisfuncs}, resembles a Gaussian function
\begin{equation}
    I_\nu=a_\nu\e^{-\frac{\theta^2}{2\sigma_h^2}}\ ,
\end{equation}
where\footnote{This expression is obtained by minimizing  
$\mathcal F(\sigma_h)=\int_0^{\theta_h}\dif\theta\theta[\frac{2\theta_h^2}{\pi\sigma_h^2}\e^{-\theta^2/(2\sigma_h^2)}-f_0\left(\frac{\theta}{\theta_h}\right)]^2$} $\sigma_h\equiv2\,\theta_h/5$. 
For a given $\sigma_h$, this Gaussian template can be used as a fake source or, in the context of a likelihood-ratio test, as the tested model.
The non-observation of diffuse signals translates into a limit on the amplitude $a_\nu$ (in Jy/beam) which \diffsph{} can reinterpret as exclusion limits on the annihilation cross section of WIMPs.

\subsubsection{Forecasts with \texttt{sigmav\_limest}} 
\label{sec:forecasts}
Another unique capability of \diffsph{} is that of producing estimated upper limits on the relevant signal-strength parameters ($\sv$, $\Gamma_\chi$, etc.) of the diffuse emission given the root-mean-square noise of the observation.
This is computed using a method devised in Ref. \cite{Vollmann:2019boa}, where it is shown that 
\begin{equation}
    \mq_\textrm{lim}\approx20.6\frac{I_\nu^\textrm{rms}\Omega_\textrm{beam}}{\sum_{n=1}^N(-1)^{n-1}h_nX_n(\nu)}\frac{2\,r_h}{\textrm{HFD}}\sqrt{\frac{\Omega(\theta_h)}{\Omega_\textrm{beam}}}\ ,
\end{equation}
where $\mq_\text{lim}$ is the estimated limit for $\mq=\sv/(2\mchi)$, $\Gamma_\chi/\mchi$, etc; HFD stands for \emph{Half-Flux Diameter} and $\Omega(\theta_h)=\pi\theta_h^2$.

\subsection{Submodules}
As depicted in Fig. \ref{fig:flowchart}, the top-level functions that are defined in the main \texttt{pyflux} and \texttt{limits} modules are constructed using the \texttt{profiles}, \texttt{spectra} and \texttt{utils} sub-modules.
We briefly describe these below

\subsubsection{Halo/bulge \texttt{profiles} module}
The \texttt{profiles} module includes halo/bulge factor ($h_n$) calculators (see Eq. \eqref{eq:hfac}) for generic $\mr(r)$ functions. We provide several benchmark templates (see Tables \ref{tab:rhos}-\ref{tab:R(r)}) for these functions as well.
These are characterized by a set of parameters, e.~g. $\rho_s$, $r_s$, etc., that can be provided by the user.
For example, the following command outputs the $h_{n=3}$ factor in GeV${}^2$/cm${}^5$ for an Einasto profile with the given parameters and assuming DM annihilation. 
\begin{table}[t!]
\centering 
\begin{tabular}{c||c|c|c}
{} & Navarro-Frenk-White & Isothermal & Hernquist-Diemand-Zhao \\
\hline
\hline\\[-5pt]
\tt rad\_temp & \tt NFW &\tt SIS &\tt HDZ \\[5pt]
$\rho(r)$ & $\displaystyle\frac{\rho_sr_s}{r(1+r/r_s)^2}$ & $\displaystyle\frac{\sigma_v^2}{2\pi G r^2}$ & $\displaystyle\frac{\rho_sr_s^\gamma}{r^\gamma(1+r^\alpha/r_s^\alpha)^{(\beta-\gamma)/\alpha}}$
\end{tabular}
\bigskip
\begin{tabular}{c||c|c|c|c|c}
{} & Einasto & Pseudo Isothermal &  Burkert &  cored NFW\\
\hline
\hline\\[-5pt]
\tt rad\_temp  & \tt Enst &\tt ps\_ISO & \tt Bkrt & \tt cNFW\\[5pt]
$\rho(r)$ & $\displaystyle\rho_s\e{}^{-\frac2{\alpha}\left(\frac{r^\alpha}{r_s^\alpha}-1\right)}$ & 
$\displaystyle\frac{\rho_s}{1+r^2/r_s^2}$ & $\displaystyle\frac{\rho_s}{(1+r/r_s)(1+r^2/r_s^2)}$ & 
$\displaystyle\frac{\rho_s}{r_c/r_s+r/r_s(1+r/r_s)^2}$ 
\end{tabular}
\caption{\label{tab:rhos} Dark Matter profile templates available in \diffsph{}.}
\end{table}

\begin{table}[!ht]
\centering 
\begin{tabular}{c||c|c|c}
{} & Plummer & Point Source & Constant \\
\hline
\hline\\[-5pt]
\tt rad\_temp & \tt Plmm &\tt PS & \tt Const \\[5pt]
$\mr_\star(r)$ & $\displaystyle\frac3{4\pi r_\star^3}\frac1{(1+r^2/r_\star^2)^{5/2}}$ & $\displaystyle\frac1{4\pi r^2}\delta(r)$ or $\displaystyle\frac1{(2\pi r_\star^2)^{3/2}}\e^{-\frac{r^2}{2r_\star^2}}$  & $\frac3{4\pi r_\star^3}$
\end{tabular}
\caption{\label{tab:R(r)} Radial function templates for generic hypotheses in \diffsph{}.}
\end{table}
\bigskip

\noindent
\begin{center}
\begin{tabular}{|ll|}
\hline \\ [-5pt]
\ttfamily \fontseries{l}\selectfont In[ ]:      &  \ttfamily {\color{jupytergreen}{\fontseries{b}\selectfont from}} {\fontseries{l}\selectfont diffsph.profiles} {\color{jupytergreen}{\fontseries{b}\selectfont import}} {\fontseries{l}\selectfont hfactors} 
\\[5pt]
{} & \ttfamily \fontseries{l}\selectfont hfactors.halo\_factor(\\
{} &  \ttfamily \fontseries{l}\selectfont\qquad n \jeq {\color{jupytergreen} 3}, rh \jeq {\color{jupytergreen} 0.8}, rad\_temp \jeq {\color{jupyterred}'enst'}, rs \jeq {\color{jupytergreen} 0.5}, rhos \jeq {\color{jupytergreen} 2.4}, hyp \jeq {\color{jupyterred}'wimp'}) \\[10pt]
\ttfamily \fontseries{l}\selectfont Out[ ]:      & \ttfamily \fontseries{l}\selectfont 3.263539652523281e+23\\[5pt]
\hline
\end{tabular}
\end{center}
\bigskip

Yet another unique feature of \diffsph{} is the outsourcing of the halo parameters, which are obtained from state-of-the-art catalogues from stellar kinematics (see Table \ref{tab:reffits}).
These can be accessed (default option) if the Boolean variable {\tt manual} is set {\tt True}.
For instance,
\bigskip

\noindent
\begin{center}
\begin{tabular}{|ll|}
\hline \\ [-5pt]
\ttfamily \fontseries{l}\selectfont In[ ]:      &  \ttfamily {\color{jupytergreen}\fontseries{b}\selectfont from} {\fontseries{l}\selectfont diffsph.profiles} {\color{jupytergreen}\fontseries{b}\selectfont import} {\fontseries{l}\selectfont massmodels} \\[10pt]
{} & \ttfamily \fontseries{l}\selectfont massmodels.h(\\
{} &  \ttfamily \fontseries{l}\selectfont\qquad n \jeq {\color{jupytergreen} 3}, galaxy \jeq {\color{jupyterred}'Ursa Minor'}, ratio \jeq {\color{jupytergreen} 0.8}, rad\_temp \jeq {\color{jupyterred}'NFW'}, \\
{} &   \ttfamily \fontseries{l}\selectfont\qquad hyp \jeq {\color{jupyterred}'wimp'}, manual \jeq {\color{jupytergreen}\fontseries{b}\selectfont False}, ref \jeq {\color{jupyterred}'Martinez'}) \\[10pt]
\ttfamily \fontseries{l}\selectfont Out[ ]:      & \ttfamily \fontseries{l}\selectfont 4.289075479711235e+23\\ [5pt]
\hline
\end{tabular}
\end{center}
\bigskip

In the previous example we introduced \texttt{ratio} variable, which is defined as the ratio $r_h/r_\star$, where $r_\star$ is the half-light radius of the galaxy.
The \texttt{profiles} module also includes brightness, emissivity and flux-density halo/bulge factor calculators in the Regime ``A'', ``B'' and ``C'' approximations introduced in Ref. \cite{Vollmann:2019boa} (see e.~g. Eqs. \eqref{eq:approxbright}, \eqref{eq:hbrightfuncs}).
In the same way as before, users can decide whether to use mass models shown in Tab.~\ref{tab:reffits} or to input the parameters manually.
Particularly interesting for prompt-emission signals (e.~g. gamma rays) are the (differential and solid-angle integrated) $J$- and $D$-factor calculators (which are equivalent to the regime ``A'' halo/bulge brightness and flux-density factors respectively).

\begin{table}[!t]
\centering 
\begin{tabular}{c||cccc}
Name & Nr. of modelled galaxies & Models & Reference \\
\hline
\hline\\[-5pt]
\tt '1309.2641' & 19 & (c)NFW, Burkert, Einasto & \cite{2015MNRAS.451.2524M}\\[5pt]
\tt '1408.0002' & 20 & HDZ & \cite{Geringer-Sameth:2014yza} \\[5pt]
\tt '1608.01749' & 1 (Draco) & HDZ & \cite{Ichikawa:2016nbi} \\[5pt]
\tt '1706.05481' & 4 & HDZ & \cite{Ichikawa:2017rph} 
\end{tabular}
\caption{\label{tab:reffits} dSph mass models linked to the \diffsph{} \texttt{profiles} module.} 
\end{table}

\subsubsection{The \texttt{spectra} module}
\label{sec:spectra}
This module allows user to compute the spectral function $X(\nu)$ (\texttt{spectra.synchrotron.X\_gen}) for a generic (user-given) source function $\ms(E)$. 
This is done by performing the following integral numerically 
\begin{equation}
    \label{eq:genspec}
    X(\nu;\tau_0)=\int_0^\infty\dif E\, \ms(E)\hat X(E;\tau_0)\ ,
\end{equation}
where $\hat X(E;\tau_0)$ is the universal spectral ``kernel'' function (see appendices). 
If the user wishes to consider a power law spectrum such as the \texttt{'generic'} entry in Table \ref{tab:modules}, they should rather use pre-computed spectral function \texttt{spectra.synchrotron.X}.

Pre-computed spectral functions for DM searches are also available. 
These are obtained from Monte Carlo parton shower simulations (in the case of DM annihilation or decay) from PPPC \cite{Cirelli:2010xx}.
\begin{equation}
\label{eq:modelspec}
    X(\nu;\tau_0) =   \frac{4\sqrt3e^3B E_\nu}{m_ec^2 b(E_\nu)}\mathcal M_{K,k\,\mchi/2}\left(\frac{2E_\nu}{k\,\mchi c^2},\eta(E_\nu);\delta\right)\quad , \quad k=1,2\ ,
\end{equation}
where $k=1$ for decaying DM and $k=2$ if WIMPs are considered. 
The function $\mathcal M_{K,M}(\xi,\eta,\delta)$ is given in the appendices. 

\subsubsection{Utilities}
The \texttt{utils} modules contains a comprehensive library of parameters that are relevant for solving the diffusion-loss equation. 
In particular all unit conversions factors and fundamental constants that are used by all modules and submodules of \diffsph{} can also be found there (see {\tt utils.consts}).

Further useful tools such as calculators of the signal's Half Flux Diameter (HFD) and Full Width at Half Maximum (FWHM) are also implemented. 
These are particularly interesting for the estimation of the limits that is mentioned in \ref{sec:forecasts} and discussed in Ref. \cite{Vollmann:2019boa}.

\section{Examples}
\label{sec:results}
\begin{figure}[ht!]
    \centering
    \includegraphics[width=.85\linewidth]{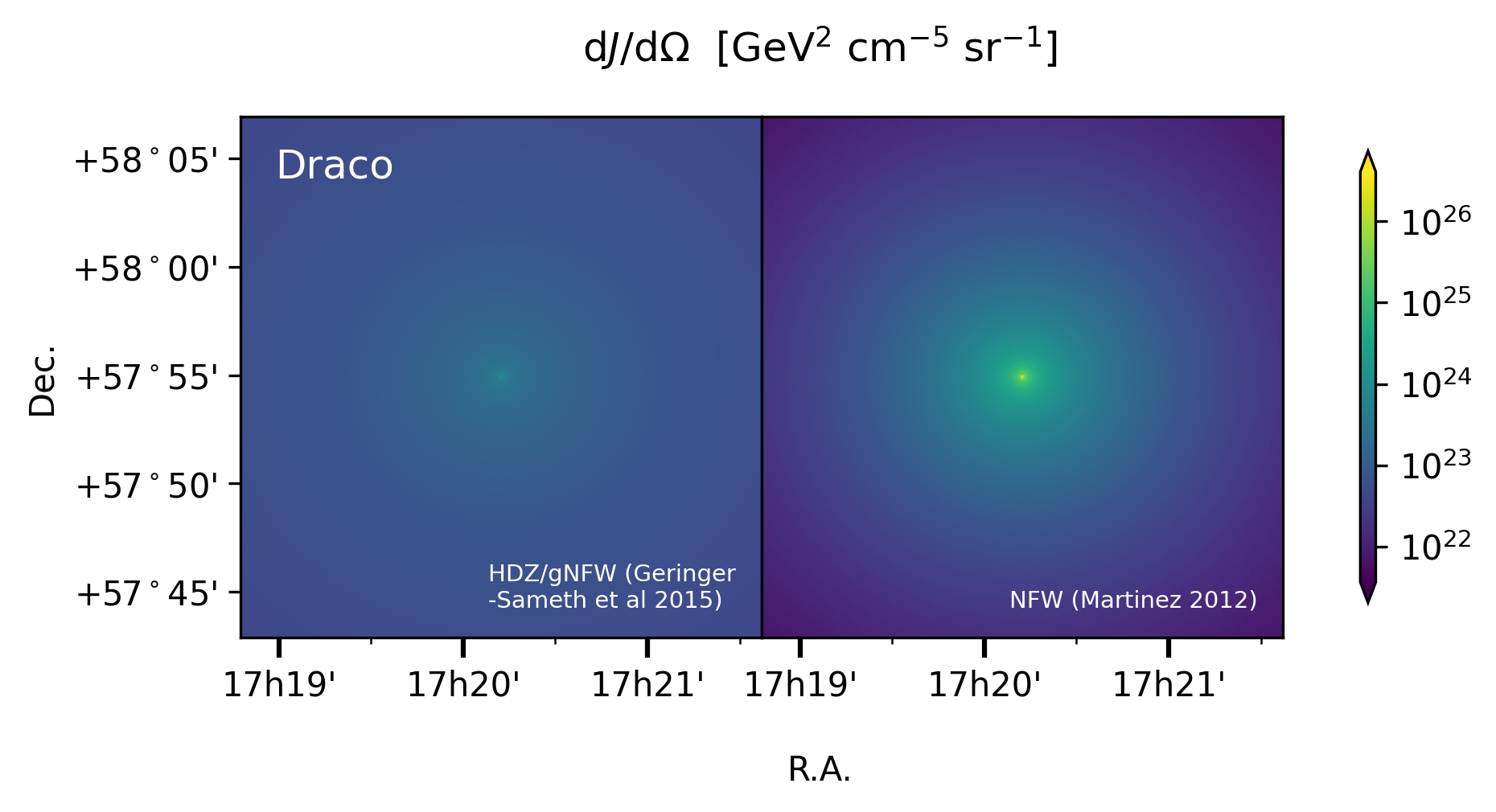}
    
    \includegraphics[width=.85\linewidth]{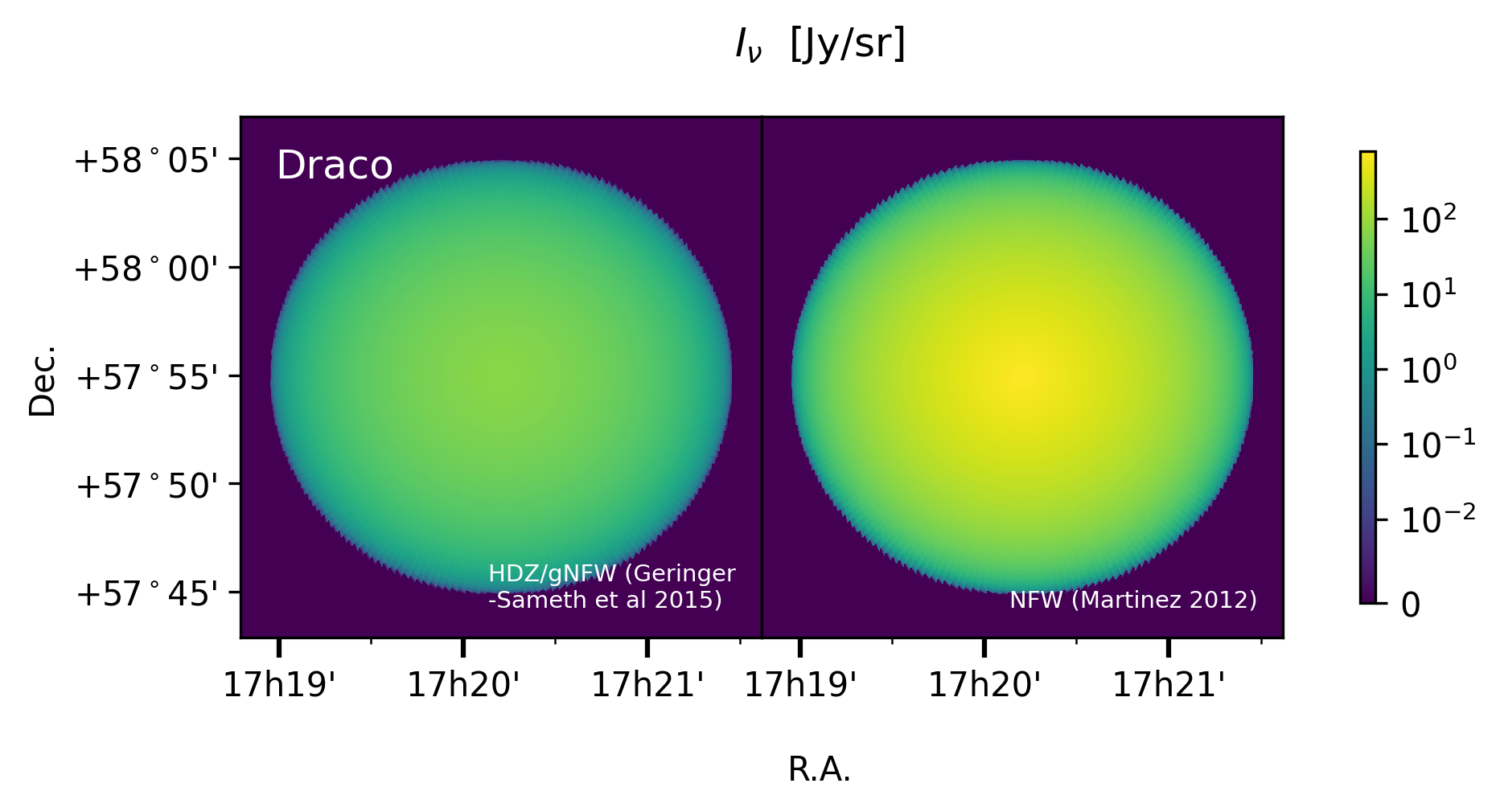}
    \caption{Diffuse emission profiles for Draco in the WIMP hypothesis for gamma rays (upper plot) and radio waves (lower plot). }
    \label{fig:profiles}
\end{figure}

The \diffsph{} tool can create signal templates for prompt (gamma rays) and synchrotron (radio) emission from DM annihilation as shown in Fig. \ref{fig:profiles}.
More specifically, by specifying a particular halo model for a given dwarf galaxy (e.~g. Draco as in the figure), users can easily generate differential $J$ and $D$ factors as functions of the galaxy's galactocentric angular distance. 
In the upper figure such $J$-factor profiles are shown. 
Two parametrizations for the DM halo are considered. 
On the left hand side the generalized NFW with the best-fit parameters from Ref. \cite{Geringer-Sameth_2015} while 
on the right hand side, the same template is plotted but using an NFW profile with model parameters extracted from Ref. \cite{2015MNRAS.451.2524M}.

In the lower panels of Fig \ref{fig:profiles} we show the synchrotron emission profile that corresponds to the aforementioned DM halo models of the Draco galaxy.
While the $J$ factors only depend on the DM halo mass function, the synchrotron maps shown in the figure as specific to the frequency of 150~MHz and a 10~GeV WIMP annihilating primarily into a $\mu^+\mu^-$ state with the default propagation parameters.

Another use case of \diffsph{} is the estimation of upper limits on the WIMP annihilation cross-section or the decay rate of decaying DM candidates given a noisy radio signal. 
\begin{figure}[ht!]
\centering
\begin{minipage}[t][][b]{.5\textwidth}
  \centering
  \includegraphics[width=.9\linewidth]{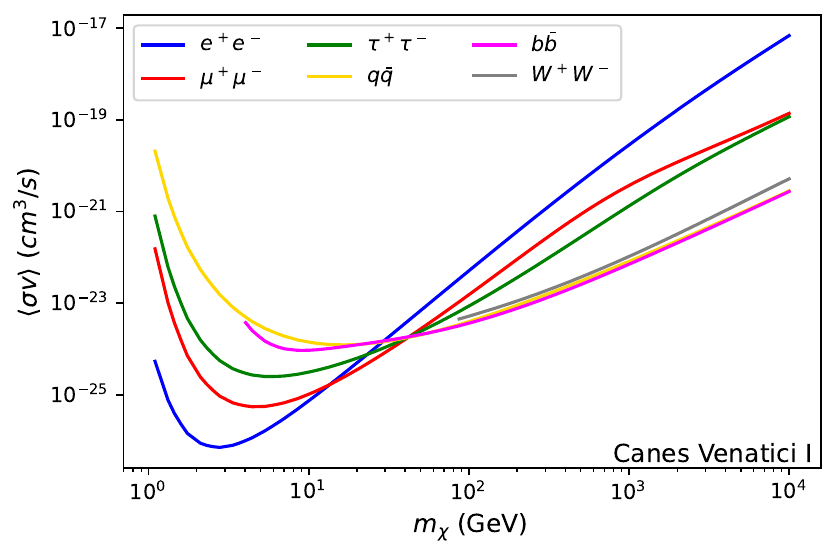}
\end{minipage}%
\begin{minipage}[t][][b]{.5\textwidth}
  \centering
  \includegraphics[width=.9\linewidth]{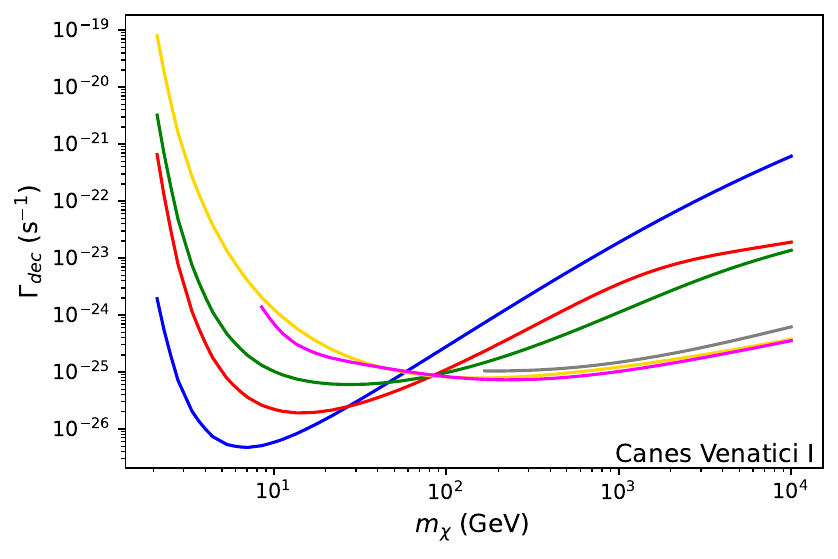}
\end{minipage}
\caption{CVnI limit estimates on the annihilation cross-section (left panel) and decay rate (right panel) of DM for several primary final states from non-observation of synchtrotron radiation.}
\label{fig:sigma_v_gamma}
\end{figure}
For that purpose, users can use the \texttt{sigmav\_limest()} and \texttt{decay\_rate\_limest()} functions contained in the \texttt{limits} module.
Fig. \ref{fig:sigma_v_gamma} shows for several primary final states the estimated upper limits that can be put on these quantities for a noisy signal of 100~$\mu$Jy/beam with a beam size of 20'' at 150~MHz.

In particular, the following example code can be used to reproduce the estimated annihilation cross-section and decay rates of DM particles as a function of their mass for $\mu^+\mu^-$ final states that shown in red in Fig. \ref{fig:sigma_v_gamma}:

\noindent
\begin{center}
\fbox{
\begin{minipage}{.91\linewidth}
\vspace{10pt}
%\setlength{\parindent}{2em}
%\noindent
\ttfamily \fontseries{l}\selectfont
{\color{jupytergreen}\fontseries{b}\selectfont from} diffsph {\color{jupytergreen}\fontseries{b}\selectfont import} limits {\color{jupytergreen}\fontseries{b}\selectfont as} lim\\
{\color{jupytergreen}\fontseries{b}\selectfont import} numpy {\color{jupytergreen}\fontseries{b}\selectfont as} np\\\\
mass\_grid \jeq np.logspace({\color{jupytergreen}0}, {\color{jupytergreen}4}) {\color{jupyterblue}\# WIMP masses}\\\\
sigmav \jeq [lim.sigmav\_limest(

\hspace{50pt} nu \jeq {\color{jupytergreen}.150}, rms\_noise \jeq {\color{jupytergreen}100}, beam\_size \jeq {\color{jupytergreen}20}, galaxy \jeq {\color{jupyterred}'Canes Venatici I'}, 

\hspace{50pt} rad\_temp \jeq {\color{jupyterred}'HDZ'}, ratio \jeq {\color{jupytergreen}1}, D0 \jeq {\color{jupytergreen}1e27}, delta \jeq {\color{jupyterred}'kol'}, B \jeq {\color{jupytergreen}1}, mchi \jeq m, 

\hspace{50pt} channel \jeq {\color{jupyterred}'mumu'}, self\_conjugate \jeq {\color{jupytergreen}\fontseries{b}\selectfont True}, manual \jeq {\color{jupytergreen}\fontseries{b}\selectfont False}, high\_res \jeq {\color{jupytergreen}\fontseries{b}\selectfont False}, 

\hspace{50pt} accuracy \jeq {\color{jupytergreen}1}, ref \jeq {\color{jupyterred}'1408.0002'}

\hspace{47pt} ) {\color{jupytergreen}\fontseries{b}\selectfont for} m {\color{jupytergreen}\fontseries{b}\selectfont in} mass\_grid] {\color{jupyterblue}\# calculating cross-sections}\\\\
Gamma \jeq [lim.decay\_rate\_limest(

\hspace{50pt} nu \jeq {\color{jupytergreen}.150}, rms\_noise \jeq {\color{jupytergreen}100}, beam\_size \jeq {\color{jupytergreen}20}, galaxy \jeq {\color{jupyterred}'Canes Venatici I'}, 

\hspace{50pt} rad\_temp \jeq {\color{jupyterred}'HDZ'}, ratio \jeq {\color{jupytergreen}1}, D0 \jeq {\color{jupytergreen}1e27}, delta \jeq {\color{jupyterred}'kol'}, B \jeq {\color{jupytergreen}1}, mchi \jeq m, 

\hspace{50pt} channel \jeq {\color{jupyterred}'mumu'}, self\_conjugate \jeq {\color{jupytergreen}\fontseries{b}\selectfont True}, manual \jeq {\color{jupytergreen}\fontseries{b}\selectfont False}, high\_res \jeq {\color{jupytergreen}\fontseries{b}\selectfont False}, 

\hspace{50pt} accuracy \jeq {\color{jupytergreen}1} ref \jeq {\color{jupyterred}'1408.0002'}

\hspace{47pt} )  {\color{jupytergreen}\fontseries{b}\selectfont for} m {\color{jupytergreen}\fontseries{b}\selectfont in} mass\_grid] {\color{jupyterblue}\# calculating decay rates}
\end{minipage}}
\end{center}

\subsection{Generic power-law spectrum}
\begin{figure}[!ht]
\begin{center}

    \includegraphics[width = .9\linewidth]{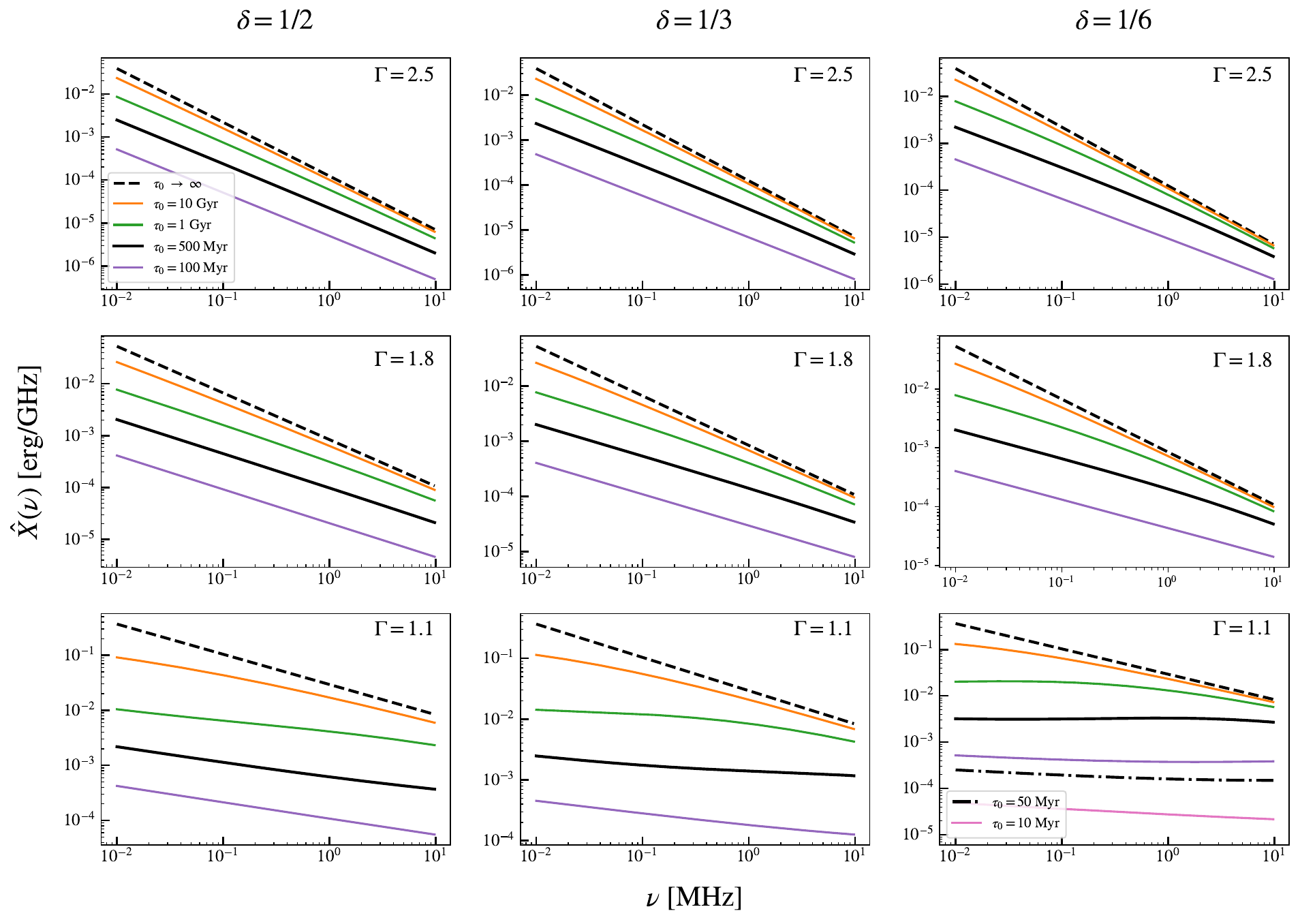}
    \caption{\label{fig:powspecs} Spectral function $X(\nu)$ for generic power-law dependent CRE source functions for several combinations of the $\Gamma$ and $\delta$ parameters. 
    }
\end{center}
\end{figure}

Another gap that is filled by \diffsph{} is considering generic (non-DM) CRE sources in dSphs. 
These have been largely neglected in the literature. 
In Fig. \ref{fig:powspecs} we show $X(\nu)$ in the case that such power-law CRE injection hypotheses are assumed.

As expected, the resulting synchrotron spectra also enjoy power-law dependences $X(\nu) \propto\nu^{-s}$. 
It is also well known that the spectral indices $s$ are related to the exponent $\Gamma$ in the CRE source function.
The way $s$ is related to $\Gamma$ is regime dependent, though.
In the fast diffusion situation (regime ``C''), the CRE distribution $n_e$ is $\propto E^{-(\Gamma+\delta)}$, which means that $s=(\Gamma+\delta-1)/2$ \cite{1979rpa..book.....R}.
In the opposite limit, where energy losses dominate, $n_e\propto E^{-(1+\Gamma)}$ and $s=\Gamma/2$.
This is precisely what is observed in all panels of the figure.

\section{Conclusions}
We introduced \diffsph{}, a computational assistant (in {\tt Python}) for the search of diffuse emissions in dwarf spheroidal galaxies.
Besides the most basic features such as computing spectra and signal profiles given some hypothesis about the CRE sources, this tool includes several additional use cases. 
In particular, the code can be used to obtain constraints on annihilating WIMPs or decaying DM models.
Non-DM scenarios, e.~g. point source with power-law CRE spectra, are also included in \diffsph{}.
The halo library is also quite comprehensive and can be used in order to compute model-specific $J$- and $D$ factors for the models contained in the dSph catalogue.

The code's excellent performance is due to a novel analytical approach that was introduced in Ref.  \cite{2020arXiv201111947V} and the incorporation and automatization of a number of built-in benchmark electron spectra (from e.~g. parton shower algorithms) and spatial distribution functions.
In future versions, modules for ICS and prompt emissions for the benchmark hypotheses presented here will also be included. 
We also aim at also enlarging the number of built-in models and improve further the code's efficiency and user experience.
\vspace{10pt}

This is an open source project licensed under an MIT License.

\section*{Acknowledgements}
LG acknowledges funding by the Deutsche Forschungsgemeinschaft (DFG, German Research Foundation) under Germany's Excellence Strategy -- EXC 2121 ``Quantum Universe'' -- 390833306. 
MV would like to thanks Kai Otto for his very useful programming hacks.

\appendix

\section{Numerical prescriptions}
\label{app:num}
In contrast to the computation of Eq. \eqref{eq:hfac}, which is rather straightforward once $\mr(r)$ is given; the spectral function \eqref{eq:spec} is much more involved and requires a dedicated treatment.
As a first step, we introduce the hypothesis independent function $\hat X(E)$, which defined in such a way that $X(\nu)$ can be expressed as
   \( X(\nu)=\int_0^\infty\dif E\, \ms(E)\hat X(E)\).
Namely,
\begin{equation}
\label{eq:defmaster}
    \hat X(E) = \frac{4\sqrt3e^3B E_\nu}{m_e c^2 b(E_\nu)}\mm\left(\frac{E_\nu}E,\eta(E_\nu),\delta\right)=
X_0\frac{B}{1+(B/B_c)^2}\frac1{E_\nu}\mm\left(\frac{E_\nu}E,\eta(E_\nu),\delta\right)\ ,
\end{equation}
where the constant $K_0$ is given by
\begin{equation}
    X_0=\frac{4\sqrt3e^3}{m_ec^2b_0}\simeq 0.0354\, \textrm{erg}\,\textrm{GHz}^{-1}\,\mu\textrm{G}^{-1}\,\textrm{GeV}\ , \quad (\texttt{utils.consts.X0}) \ .
\end{equation}

The ``master'' function $\mm$ is defined in terms of the pitch-angle averaged synchrotron-radiation function $\mf(y)$ as follows
\begin{equation}
    \label{eq:master}
    \mm(\xi,\eta;\delta)=\int_\xi^\infty{\rm d}z\,\mf(z^2)\exp\left[-\eta\left(z^{1-\delta}-\xi^{1-\delta}\right)\right] \ ,
\end{equation}
where the $\mf$ function reads
\begin{equation}
    \label{eq:fav}
    \mf(y)=y^2\left(K_{4/3}(y)K_{1/3}(y)-\frac35\left[K_{4/3}^2(y)-K_{1/3}^2(y)\right]\right) \ ,
\end{equation}
and $K_{1/3}(y)$, $K_{4/3}(y)$ are the modified Bessel functions with indices 1/3 and 4/3 respectively.
As evident by Eq. \eqref{eq:defmaster}, once the master function is known for all (positive) values of $\xi$ and $\eta$ (and, of course, $\delta$) we can determine $\hat X$ for generic parameter sets. 
Nevertheless, the kernel $\mm$ is still a very complicated function. 
Its numerical computation can take very long and for some choices of the parameters, standard integration algorithms fail to converge.
In order to overcome this, the following approximations are very useful and have been implemented in \diffsph{}.  
\begin{eqnarray*}
    M(\xi,\eta,\delta) &\to& \int_\xi^\infty{\rm d}x\,\mf(x^2) \quad , \ \eta\to0\quad (\eta\ll \xi^{-(1-\delta)})\\
    M(\xi,\eta,\delta) &\to& \int_\xi^\infty{\rm d}x\,f_0\times x^{2/3}\exp[-\eta(x^{1-\delta}-\xi^{1-\delta})]=f_0\Gamma\left(\frac5{3(1-\delta)}\right)\frac{\eta^{-\frac5{3(1-\delta)}}{\rm e}^{\eta \xi^{1-\delta}}}{1-     \delta}\quad ,\\
    {} &{}&{} \ \eta\to\infty\quad (\eta\gg (1/3)^{-(1-\delta)}\quad, \quad \eta^{1-\delta}\ll (1/3)^{1-\delta})\\
    M(\xi,\eta,\delta) &\to& \int_u^\infty{\rm d}x\,\mf(x^2)\delta(\eta(x^{1-\delta}-\xi^{1-\delta}))=\frac{\xi^\delta}{(1-\delta)\eta}\mf(\xi^2) \quad , \ \eta\to\infty\quad (\eta\gg \xi^{-(1-\delta)})            
\end{eqnarray*}

In practice, however, the way \diffsph{} evaluates $\mm$ is by interpolating pre-computed values in the logarithmic spaced grid $\log\xi\in[-15,+1.2]$, $\log\eta\in[-3,+13]$ and for selected values of the $\delta$ parameter: $\delta = 1/6$, $2/6$ ({\tt 'kol'}), $3/6$ ({\tt 'kra'}) and $4/6$, where {\tt 'kol'} and {\tt 'kra'} stand for the standard Kolmogorov \cite{1941DoSSR..30..301K} and Kraichnan \cite{Kraichnan:1965zz} CR propagation scenarions respectively.
The user can certainly turn this option off by setting the Boolean \texttt{'fast\_comp'} variable in ({\tt spectra.synchrotron.htX()}) to 0.
The ranges were chosen in a way that the extrapolations behave as expected from known asymptotic expressions. 

\subsection{Spectra from parton showers}
The functions $\mm$ and $\hat X$ are by construction hypothesis independent. 
Therefore, \diffsph{} users can compute synchrotron spectra for basically any ansatz for $\ms(E)$ by using Eq. \eqref{eq:genspec} (see {\tt spectra.synchrotron.X\_gen()}).
However, for those users interested in testing specific particle DM scenarios in which the DM annihilates or decays primarily into two-body standard-model final states ($e^+e^-, b\bar b,\ldots$), \diffsph{} contains a number of pre-computed spectral functions $X_{k;K,\mchi}(\nu)$, where $K=e^+e^-, b\bar b,\ldots$ and the index $k$ specifies the process: $k=1$ for decay and $k=2$ for annihilation. 
These are obtained using the analogue of Eq. \eqref{eq:defmaster}, where instead of the master function $\mm$ we use 
\begin{equation}
\label{eq:MSa}
    \mathcal M_{K,k\,\mchi/2}(\xi,\eta;\delta) = \int_0^1{\rm d}x\,\mm\left(\frac{\xi}x,\eta,\delta\right)\,s_{K,k\,\mchi/2}(x)\quad , \quad k=1,2\ ,
\end{equation}
where $x=E/(k\,\mchi/2)$, $s_{K,k\,\mchi/2}(x)=(k\,\mchi/2)\times\ms_{K,k\,\mchi/2}(E)$ and $\ms_{K,k\,\mchi/2}(E)$ is the injection function which is obtained from parton shower algorithm assuming a two-body final state $K$. 

\subsection{Spectra from power-law CRE distribution}
For a generic power-law CRE distribution $\ms^0(E)=E_0^{\Gamma-1}/E^\Gamma$, we define
\begin{eqnarray}
\nonumber
    \mathcal M_{\Gamma}(\xi,\eta,\delta) &=& \int_0^\infty\frac{\dif x}{x^\Gamma}\,\mm\left(\frac{\xi}x,\eta,\delta\right)=\xi^{-(\Gamma-1)}\int_0^\infty\dif \xi'\,\xi'^{\Gamma-2}\mm\left(\xi',\eta,\delta\right)\ ,\\
 \label{eq:MSa0}
    {} &=& \frac1{\xi^{\Gamma-1}}\int_0^\infty \dif z\,\mf(z^2)\e{}^{-\eta\, z^{1-\delta}}f_0(z;\eta,\delta) \ ,
\end{eqnarray}
where
\(f_0(z;\eta,\delta)=\int_0^z\dif\xi\,\xi^{\Gamma-2}\e{}^{\eta\,\xi^{1-\delta}}\)
can be obtained analytically.

\section{Analytic limits}
Depending on how large or small the propagation parameter $\tau_0$ is with respect to the typical energy-loss time scale $t_{\rm loss}\simeq E_{\rm max}/b(E_{\rm max})$, one can neglect either the diffusion-coefficient term (regime ``A'') or all energy losses (regime ``C'')  in Eq. \eqref{eq:diffloss} \cite{2020arXiv201111947V}. 
In the intermediate case, where none of the terms in the diffusion-loss equation can be neglected, 
one can make the approximation that all terms with $n>1$ in Eq. \eqref{eq:bright} are negligible.
Regardless of the regime, Eq. \eqref{eq:bright} can be further simplified to
\begin{equation}
\label{eq:approxbright}
I_\nu^a\approx\frac{\mq^a}{4\pi}\,\hat H_{\rm reg.}^a(\theta)\times X^a_{n=1}(\nu) \ ,
\end{equation}
where 
\begin{equation}
\label{eq:hbrightfuncs}
\hat H_{\rm RA}^a(\theta) = \sum_{n=1}^\infty \,h^a_n f_n\left(\frac\theta{\theta_h}\right)\ , \ \hat H_{\rm RB}^a(\theta) = h^a_1 f_1\left(\frac\theta{\theta_h}\right)\ {\rm and}\ 
\hat H_{\rm RC}^a(\theta) = \sum_{n=1}^\infty \frac1{n^2}\,h^a_n f_n\left(\frac\theta{\theta_h}\right)\ .
\end{equation}

Alternative (but equivalent) expressions can be found in Ref. \cite{2020arXiv201111947V}.
These expressions are implemented in the {\tt profiles} module.

\end{document}